# On the Robustness of the Delay-Based Fingerprint Embedding Scheme

Shiguo Lian

*France Telecom R&D Beijing*
*2 Science Institute South Rd., Beijing 100080, China*

**Abstract**

The delay-based fingerprint embedding was recently proposed to support more users in secure media distribution scenario. In this embedding scheme, some users are assigned the same fingerprint code with only different embedding delay. The algorithm's robustness against collusion attacks is investigated. However, its robustness against common desynchronization attacks, e.g., cropping and time shifting, is not considered. In this paper, desynchronization attacks are used to break the delay-based fingerprint embedding algorithm. To improve the robustness, two means are proposed to keep the embedded fingerprint codes synchronized, i.e., adding a synchronization fingerprint and adopting the relative delay to detect users. Analyses and experiments are given to show the improvements.

## 1. INTRODUCTION

With the development of multimedia technology and network technology, multimedia distribution becomes more and more popular. In some applications, such as video-on-demand, broadcasting and IPTV, some authorized users may redistribute their media copy to other unauthorized users, which cause profit-losses to service providers. As a solution, digital fingerprinting [1] was recently reported, which embeds users' identification information into media content imperceptibly and produces different copies for different users. The identification information that is often named fingerprint code can be used to trace illegal users.

However, fingerprinting techniques face a big threat named collusion attack [1]. In collusion attack, several users combine their copies together to generate a new copy without fingerprint codes. The combination operation may be averaging, min-max selection, linear combination, etc. To resist collusion attacks, some means have been proposed to generate fingerprint codes, which can be classified into two types, i.e., orthogonal fingerprint and combinatorial code. In orthogonal fingerprint [2][3][4], the fingerprint code is the sequence independent from each other. For example, the fingerprint code can be a bounded Gaussian (BG), pseudo-noise (PN) or orthogonal Hadamard Walsh codes, and different fingerprint code corresponds to different user. In combinatorial code, fingerprint codes are carefully designed by combinatorial codeword [1,5,6], which can detect the colluders partially or completely. The combinatorial codes have the following property: each group of colluders' fingerprint produces unique codeword that determines all the colluders in the group. The codeword is constructed based on combinatorial theory, such as AND-ACC (anti-collusion codes) or BIBD [7].

According to the existing fingerprint schemes, different user has a different fingerprint code. Thus, for the scheme supporting N users, N fingerprint codes should be generated and stored. To reduce the time and space cost, the delay-based fingerprint embedding scheme [8] is recently reported. In this scheme, the users are classified into groups, and the users in the same group have the same fingerprint code, with only the embedding delay different. Thus, the number of fingerprint codes can be reduced greatly. The scheme's robustness against collusion attacks has been analyzed and confirmed. However, the robustness against such desynchronization attacks as cropping and time shifting has not been considered, which restricts its application greatly.

In this paper, the delay-based fingerprint embedding scheme's robustness against desynchronization attack is analyzed and tested. Based on the analyses, some means are presented to improve the scheme's robustness. And the improvements will be shown by the experiments. The rest of the paper is arranged as follows. In Section 2, the delay-based fingerprint embedding scheme is introduced. The desynchronization attacks are used to break the delay-based embedding scheme in Section 3. In Section 4, some means are presented to improve the scheme's robustness, and the experiments are shown in Section 5. Finally, some conclusions are drawn and future work is given in Section 6.

## 2. THE DELAY-BASED FINGERPRINT EMBEDDING SCHEME

In traditional fingerprinting algorithms [1], each user is assigned a different fingerprint code. Thus, for N users, N fingerprint codes should be generated and stored. In the delay-based fingerprint embedding scheme [8], as shown in Fig. 1, *N* users are partitioned into *M* groups, each of which is composed of *P (N=MP)* users. For the P users in the same group, i.e., *i*-th *(i=0,1,…,M-1)* group, they have the same fingerprint code $F_i$ but different embedding delay $d_j$ *(j=0,1,…,P-1)*. Thus, the number of fingerprint codes is reduced from *N* to *M*.

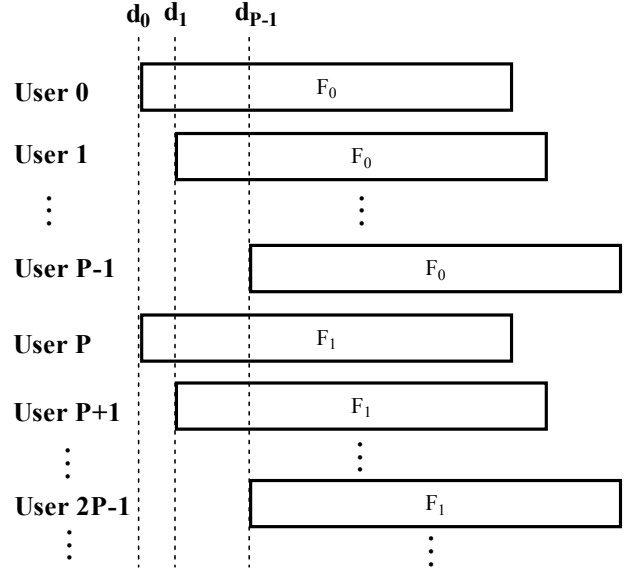

**Fig. 1.** The delay-based embedding model

Generally, the fingerprint code is embedded into media content repeatedly in order to improve the robustness against common signal processing. Set $X=x_0x_1...x_{n-1}$ the original content segment, $\alpha$ the embedding strength, and $Y_t=y_{t,0}y_{t,1}...y_{t,n-1}$ the t-th user's fingerprinted content segment. For the *t*-th *(t=0,1,...,N-1)* user, his fingerprint code is $F_i=w_{i,0}w_{i,1}...w_{i,n-1}$ ($i=\lfloor t/P \rfloor$, and $\lfloor t/P \rfloor$ is the maximal integer no bigger than *t/P*), and embedding delay is $d_j$ (*j=t* mod *P*, and mod is module operation). Taking additive embedding for example, the *t*-th embedding is defined as

$$y_{t,k-d_j} = x_{k-d_j} + \alpha \,|\, x_{k-d_j}\,|\, w_{i,k} \quad (k=0,1,\cdots,n-1). \tag{1}$$

The correlation based detection is used to detect the existence of the fingerprint codes $F_i$ and the corresponding delays $d_j$, see [8] for detail. It has been show that the delay-based algorithm saves some cost for fingerprint embedding/detection, and it is robust against some collusion attacks. However, the robustness against such desynchronization attack as cropping or time shifting is not confirmed, which will be discussed as follows.

## 3. ATTACK ON THE DELAY-BASED SCHEME

### 3.1 Desynchronization Attacks

Desynchronization attacks are often used to break watermarking systems, which change the watermarked media's position in spatial domain or temporal domain in order to make the watermark undetectable. Taking audio watermarking for example, the typical desynchronization attacks include crop-ping, time shifting, etc. As shown in Fig. 2, cropping is to cut off a block from the watermarked audio, while time shifting is to change the time position of audio content.

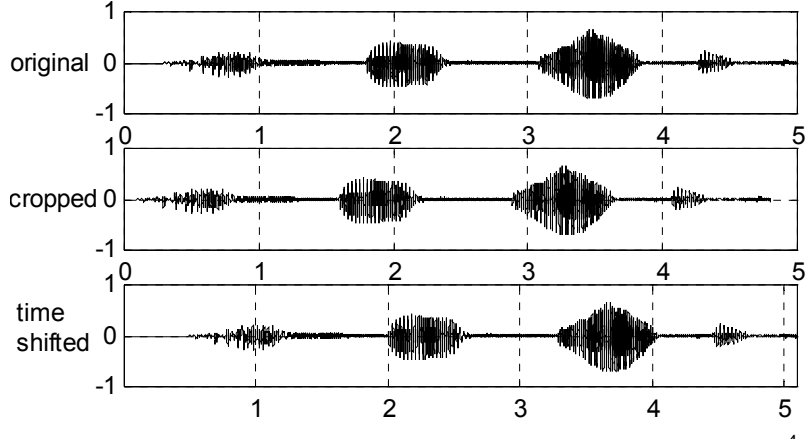

**Fig. 2**. Examples of cropping and time shifting

**3.2 Desynchronization Attack without Collusion**

The desynchronization attacks can be applied to the watermarked audio content when there are no collusion attacks. Taking the audio content contains the t-th user's fingerprint code $F_i$ and the corresponding embedding delay $d_j$ for example, as shown in Fig. 3, the cropping or time shifting attacks can easily change the delay. Here, after cropping or shifting the watermarked audio content, the fingerprint's position is moved from $d_j$ to $d_{j'}$ or $d_{j''}$.

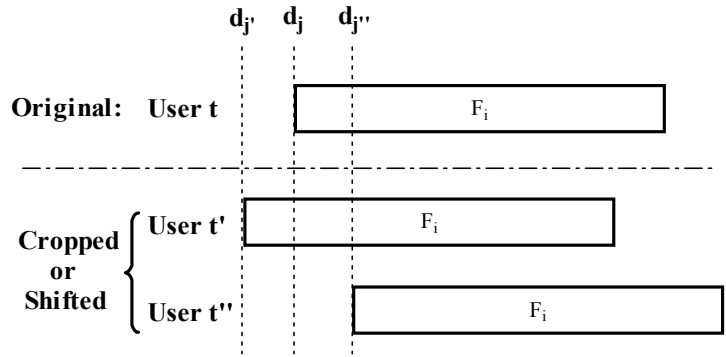

**Fig. 3.** Desynchronization attacks without collusion

Taking the case of $d_{j'}$ for example, the attacked audio content $Y_{t'}=y_{t',0}y_{t',1}...y_{t',n-1}$ satisfies

$$y_{t',k-d_{j'}} = x_{k-d_{j'}} + \alpha \mid x_{k-d_{j'}} \mid w_{i,k} \quad (k=0,1,\cdots,n-1). \tag{2}$$

Thus, after detecting the fingerprint code $F_i$ and the delay $d_{j'}$, the user $t'$ can be determined, which satisfies

$$t' = i \cdot P + j'. \tag{3}$$

Then, the detected user is $t'$. Since $j' \neq j$, then $t' \neq t$. Thus, after desynchronization attacks, it is difficult to detect the user correctly.

### 3.3 Desynchronization Attack after Collusion

After collusion attacks, desynchronization attacks can be fatherly used to change the watermarked audio content. As shown in Fig. 4, for two colluders, $t_0$ and $t_1$, their fingerprint codes are $F_{i0}$ and $F_{i1}$, and their embedding delays are $d_{j0}$ and $d_{j1}$, respectively. After cropping or time shifting, the fingerprint codes' positions may be changed from $d_{j0}$, $d_{j1}$ to $d_{j'0}$, $d_{j'1}$ or $d_{j''0}$, $d_{j''1}$.

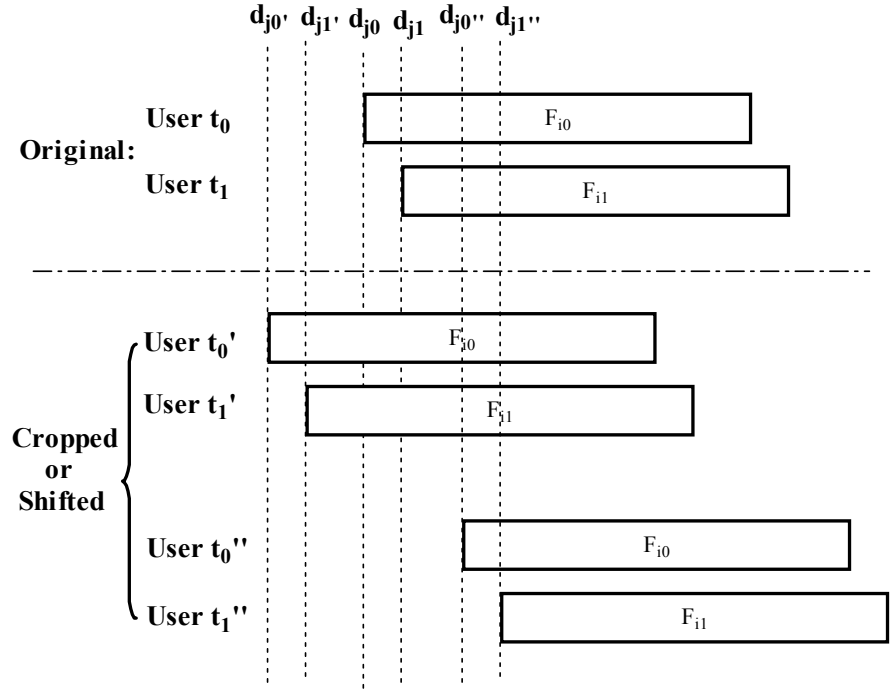

**Fig. 4.** Desynchronization attacks after collusion

Taking the case, in which, the embedding delays are $d_{j'0}$ and $d_{j'1}$, for example, the attacked audio contents $Y_{t'0} = y_{t'0,0} y_{t'0,1} \ldots y_{t'0,n-1}$ and $Y_{t'1} = y_{t'1,0} y_{t'1,1} \ldots y_{t'1,n-1}$ satisfy

$$\begin{cases} y_{t'_0, k-d_{j'0}} = x_{k-d_{j'0}} + \alpha \, |x_{k-d_{j'0}}| \, w_{i_0, k} \\ y_{t'_1, k-d_{j'1}} = x_{k-d_{j'1}} + \alpha \, |x_{k-d_{j'1}}| \, w_{i_1, k} \end{cases} \tag{4}$$

Here, k=0,1,…,n-1. Thus, after detecting the fingerprint codes, $F_{i0}$ and $F_{i1}$, and the delays, $d_{j'0}$ and $d_{j'1}$, the colluders, $t'_0$ and $t'_1$, can be determined by

$$\begin{cases} t_0' = i_0 \cdot P + j_0' \\ t_1' = i_1 \cdot P + j_1' \end{cases}. \qquad (5)$$

Since $j_0' \neq j_0$ and $j_1' \neq j_1$, then $t_0' \neq t_0$ and $t_1' \neq t_1$. Thus, after desynchronization attacks, the colluders can not be detected correctly.

## 4. MEANS TO IMPROVE THE DELAY-BASED FINGERPRINT EMBEDDING SCHEME

According to the analyses presented in Section 3, the delay-based fingerprint embedding scheme is not robust against desynchronization attack. Here, two means are proposed to improve the scheme's robustness: i) add an additional synchronization fingerprint, and ii) detect the fingerprint with comparative delay. The fingerprint codes and the corresponding delays are shown in Fig. 5.

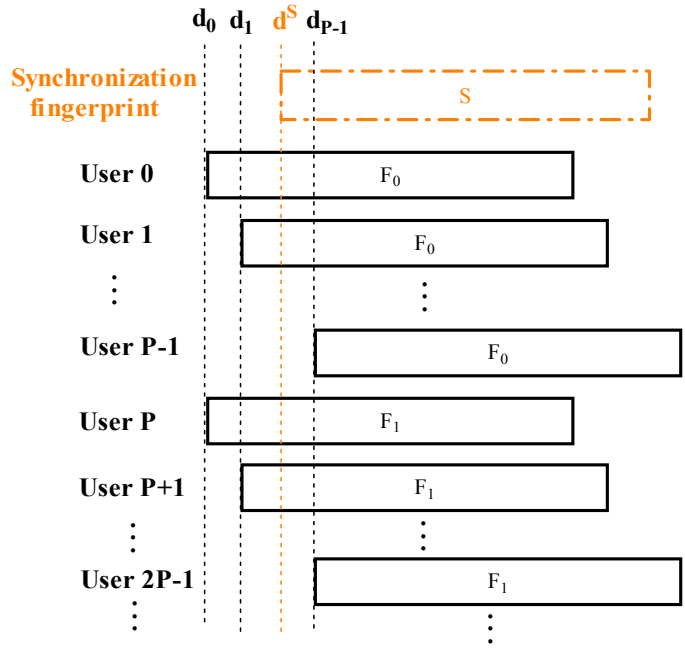

**Fig. 5.** The improved delay-based embedding model

In the improved scheme, for the t-th user, the synchronization fingerprint $S=s_0s_1...s_{n-1}$ are embedded into the media content besides the fingerprint $F_i$. Here, $S$ is orthogonal to $F_i$. The embedding process is defined as

$$y_{t,k-d_j} = x_{k-d_j} + \alpha |x_{k-d_j}| w_{i,k} + \alpha |x_{k-d^S}| s_k. \qquad (6)$$

Here, *k=0,1,…,n-1*, $d^S$ is the embedding delay of the synchronization fingerprint, and other parameters are similar to the ones defined in Eq. (1). The delay $d^S$ can be randomly selected, but it keeps the same for different customer.

The correlation based detection is used to detect the position of synchronization fingerprint, $d^S$, the existence of the fingerprint codes $F_i$ and the corresponding delays $d_j$. If no desynchronization attacks exist, then, the t-th user can be detected by

$$t = i \cdot P + j. \tag{7}$$

And the colluders are determined by $d_j$-$d^S$.

If desynchronization attacks exist, the improved scheme uses the comparative delays to remove the effect caused by desynchronization attacks. Taking the desynchronization attacks without collusion for example, after attacks, the detected delay of *S* is $d^{S'}$, and the detected delay of $F_i$ is $d_{j'}$. Then, the original $d_j$ can be computed by

$$d_j = d_{j'} - (d^{s'} - d^s). \tag{8}$$

And then, the *t*-th user can be correctly detected according to Eq. (7). Under the condition of desynchronization attacks with collusion, the similar result can be induced.

## 5. PERFORMANCE COMPARISON

Taking the popular music Casblk1.wav sampled at 44.1KHz for example, the robustness of the original delay-based scheme and the improved scheme are compared. Here, P=4, M=16, N=64, n=1024, $\alpha = 0.05$, $d_{j+1}$-$d_j$=20, and $d^S$ is randomly generated. For simplicity, the synchronization fingerprint *S* and the fingerprint code $F_i$ are generated by pseudo-noise [8] with the maximal amplitude of 1. In the desynchronization attacks, each segment composed of 1024 pixels is cropped or time-shifted in the range of [1,512]. Under the condition of desynchronization without collusion, 100 copies are generated for the original scheme and the improved scheme, respectively. In the 100 copies, there are 50 cropped copies and 50 shifted copies. The correct detection rate is computed for different schemes, which denotes the ratio between the number of correctly-detected copies and the total number of copies. For the original scheme, the correct detection rate is no more than 25%, while, for the improved scheme, the correct detection rate is 100%. Under the condition of desynchronization with collusion, the average-colluded copies are cropped or shifted, and 100 copies are also produced. The comparison between the original scheme and the improved scheme is shown in Fig. 6. Seen from it, the improved scheme gets higher correct detection rate than the original scheme, and obtains the similar result with the case without desynchronization attack. It shows that the improved scheme is much more robust against desynchronization attacks than the original scheme.

The improved scheme's robustness against some other attacks is similar to the original scheme. The original scheme's robustness has been analyzed and evaluated in [8][9]. In the improved scheme, two fingerprint codes are embedded into each media copy. Compared with the original scheme, the improved one causes some quality degradation. However, it has been shown in [9] that the degradation is imperceptible when the number of embedded fingerprint codes is no more than 4. Thus, in the improved scheme, the fingerprint codes are still imperceptible.

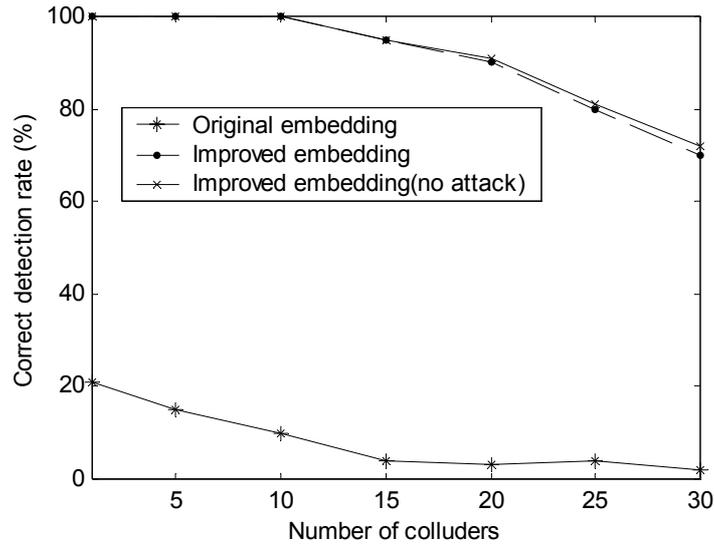

**Fig. 6.** The robustness against desynchronization attack

## 6. CONCLUSIONS AND FUTURE WORK

In this paper, the robustness of the delay-based fingerprint embedding scheme is analyzed and improved. Firstly, desynchronization attacks, including cropping and time shifting, are used to break the delay-based fingerprint scheme. It is shown that the colluders can easily escape from the detection if the fingerprinted media is desynchronized. Following the attacks, an improved scheme is proposed, which adds a synchronization fingerprint and adopts the comparative embedding delay to trace colluders. Analyses and experiments show that the improved scheme obtains much higher robustness compared with the original scheme, while its other performances keep nearly unchanged. In future work, the scheme's robustness against some other attacks will be investigated, and the fingerprint code with better robustness will be evaluated.